\documentclass[%
 reprint,
 amsmath,amssymb,
 aps,
nofootinbib
]{revtex4-2}

\usepackage{graphicx}% Include figure files
\usepackage{dcolumn}% Align table columns on decimal point
\usepackage{bm}% bold math
\usepackage[mathlines]{lineno}% Enable numbering of text and display math
\usepackage{color}
\def\D{\mathrm{D}}
\def\dd{\mathrm{d}}

\begin{document}

%\preprint{AAPM/123-QED}

\title{Time-dependent $G$ in Einstein's equations as an alternative to the cosmological constant
}% Force line breaks with \\

\author{Ekim Taylan Han{\i}meli}
 \email{ekimtaylan@gmail.com}
 \affiliation{Universit{\'e} de Toulouse, UPS-OMP, IRAP, CNRS, 14 Avenue Edouard Belin, F-31400 Toulouse, France}%Lines break automatically or can be forced with \\

\author{Brahim Lamine}
%\email{}
\affiliation{Universit{\'e} de Toulouse, UPS-OMP, IRAP, CNRS, 14 Avenue Edouard Belin, F-31400 Toulouse, France}

\author{Isaac Tutusaus}
%\email{}
\affiliation{Institute of Space Sciences (ICE, CSIC), Campus UAB, Carrer de Can Magrans, s/n, 08193 Barcelona, Spain}
\affiliation{Institut d`Estudis Espacials de Catalunya (IEEC), Carrer Gran Capit\`a 2-4, 08193 Barcelona, Spain}
\affiliation{Universit{\'e} de Toulouse, UPS-OMP, IRAP, CNRS, 14 Avenue Edouard Belin, F-31400 Toulouse, France}

\author{Alain Blanchard}
%\email{}
\affiliation{Universit{\'e} de Toulouse, UPS-OMP, IRAP, CNRS, 14 Avenue Edouard Belin, F-31400 Toulouse, France}

\date{\today}% It is always \today, today,

\begin{abstract}
In this work, we investigate cosmologies where the gravitational constant varies in time, with the aim of explaining the accelerated expansion without a cosmological constant. We achieve this by considering a phenomenological extension to general relativity, modifying Einstein's field equations such that $G$ is a function of time, $G(t)$, and we preserve the geometrical consistency (Bianchi identity) together with the usual conservation of energy by introducing a new tensor field to the equations. In order to have concrete expressions to compare with cosmological data, we posit additional properties to this tensor field, in a way that it can be interpreted as a response of spacetime to a variation of $G$. Namely, we require that the energy this tensor represents is nonzero only when there is a time variation of $G$, and its energy depends on the scale factor only because of its coupling to $G$ and the matter and radiation energy densities. Focusing on the accelerated expansion period, we use type Ia supernovae and baryon acoustic oscillation data to determine the best fit of the cosmological parameters as well as the required variation in the gravitational constant. As a result, we find that it is possible to explain the accelerated expansion of the Universe with a variation of $G$ and no cosmological constant. The obtained variation of $G$ stays under 10 \% of its current value in the investigated redshift range and it is consistent with the local observations of $\dot{G}/G$.

\end{abstract}

\maketitle

\section{Introduction}
The concordance model in cosmology, $\Lambda$CDM, is extremely successful in being able to explain most of the current cosmological observations with great precision~\cite{obsprobes, Blanchard2010, planck, boss,pantheon,DEC}. On the other hand, this model also has important problems, one of which is its inability to explain the nature of the titular $\Lambda$, or the cosmological constant, which remains to be an \textit{ad hoc} addition to general relativity, employed in order to explain the late stage accelerated expansion of the Universe. While this constant behaves in the same way as a vacuum energy, its value is many orders of magnitude smaller than the estimations of quantum field theory~\cite{weinberg}. Finding a cosmological solution with $\Lambda=0$ solves one aspect of this problem, since this solution would be compatible with a renormalization to zero, a more natural value compared to the current estimations of $\Lambda$~\cite{beyondlcdm}.

In this paper, we present an alternative picture, in which the accelerated expansion attributed to the cosmological constant appears naturally as a result of a variation of $G$ in a relativistic model of gravitation. We achieve this by positing a phenomenological time variation of the gravitational constant $G$ in Einstein's field equations. As is well-known~\cite{barrow,sultana}, and as we will later demonstrate more clearly, this scenario creates a coupling between the matter and radiation energy density and $G$, which breaks the energy conservation of matter and radiation. We solve this issue in a general way by adding a new dynamical term in Einstein's equations. Our approach here is similar to Jordan-Brans-Dicke (JBD) theories~\cite{jordan,bransdicke}, in that we decouple the density evolution of the matter and radiation from the variation in the gravitational constant. However, we determine the gravitational constant phenomenologically from the observations, instead of obtaining it according to first principles (unlike the JBD Lagrangian, which is obtained from Mach's principle).
We focus on the accelerated expansion period and show that even a small variation of the gravitational constant can produce a similar effect to a cosmological constant in the present epoch.

In order to obtain specific cosmological models to test against the data, we make additional physical assumptions about the new dynamical term. We consider this tensor to represent only the reaction of the spacetime geometry to any variation of the gravitational constant, such that it can be interpreted as a property of the spacetime itself. Therefore, we require (a) that, apart from its dependence on other terms, the energy represented by this tensor should not be directly affected by the expansion of the space
and (b) that only the terms coupled to the evolution of the gravitational constant should be present. After this, we propose a Taylor expansion for the gravitational function $G$ around today (scale factor $a_0=1$), taking only the first few terms.

We then confront this picture with the observations of type Ia supernovae (SNIa) and baryon acoustic oscillations (BAO). While treating SNIa data, we also take into account the effect that a variation of $G$ might have on SNIa intrinsic luminosity, employing the approach widely used in the cosmology literature~\cite{gaztanaga,garciaberro,Riazuelo,Kazantzidis,mould_2014}, which assumes the supernovae intrinsic luminosity to be proportional to the Chandrasekhar mass. 
Additionally, we use the results of a more recent analysis by Ref.~\cite{wrightli}, which provides an opposite luminosity-$G$ relation, in order to see to what extent our results are affected by the physics of SNIa. 
On the other hand, we do not modify BAO data under the assumption that they will not be considerably affected by a small variation of the gravitational constant. As a result, we show that the variation of $G$ required to fit these observations is small enough to be compatible with the local constraints of $\dot{G}/G$~\cite{Uzan}.

The paper is organized as follows: in Sec.\,\ref{sec2}, we specify the problem that arises when $G$ evolves as a function of time 
and detail our general solution. In Sec.\,\ref{sec4}, we obtain the cosmological model, together with a specific $G$ function to test against the data. After detailing the data and the methodology used in Sec.\,\ref{sec5}, we present the results in Sec.\,\ref{sec6} and conclude in Sec.\,\ref{sec7}.

\section{Field equations with varying $G$}\label{sec2}

In this work, we take a phenomenological approach to modeling the variation of the gravitational constant. More explicitly, instead of determining how $G$ should behave from some starting assumptions (such as Mach's principle), and deriving our equations from an appropriate Lagrangian, we treat it as a free parameter in Einstein's field equations, whose value is to be determined by observations. This approach of letting the data choose the preferred variation of $G$ allows us to be more general in terms of assumptions beyond general relativity. In other words, rather than offering a new theory of gravity, we stick to general relativity and extend it phenomenologically by allowing G to vary in time.

Let us, then, consider general relativity with a time-dependent gravitational constant. We start with modified field equations such that
\begin{equation}
G^{\mu \nu}(\bm{x},t)= 8\pi G(\bm{x},t) T^{\mu \nu}(\bm{x},t)\,.
\label{eq:1}
\end{equation}

This equation is local; namely, it relates the Einstein tensor at a given spacetime event to the energy-momentum tensor and the gravitational constant at the same event. In general, this need not be the case, and we find in the literature various nonlocal theories generalizing Einstein's equations by incorporating retardation effects through a susceptibility function~\cite{Belgacem2018}. However, any nonlocal approach behave as being quasilocal when the susceptibility is very stitched around zero. In this situation, the response time of spacetime itself is supposed to be very small compared to other characteristic times (here, a cosmological timescale).

In accordance with the cosmological principle, we consider only a time dependence of $G$ in Eq.~(\ref{eq:1}). In this case, the Bianchi identity implies a nonzero covariant derivative for the stress energy tensor:
\begin{equation}
\D_\mu G^{\mu\nu} = 0 \quad\Rightarrow\quad\D_\mu T^{\mu\nu} = -\frac{T^{\mu \nu} \partial_\mu G(t)}{G(t)} \not = 0\,.
\end{equation}

Therefore, without any other prescription, the energy-momentum will no longer be conserved. More precisely, if one assumes the standard form for a cosmological perfect fluid, $T^{\mu \nu}=(\rho +p)u^\mu u^\nu + pg^{\mu \nu}$, where $u^\mu = (1,0,0,0)$ is the Hubble flow and $p$ and $\rho$ are the pressure and energy densities, respectively, for the usual matter and radiation, one obtains
\begin{equation}
\label{eq:non cons energy}
\dot{\rho} +3 H(\rho + p)=-\rho\frac{\dot{G}}{G}\,,
\end{equation}
where $H=\dot{a}/a$ is the Hubble parameter. The source term in the right-hand side creates a coupling between the energy density and $G$. For instance, one obtains for nonrelativistic matter ($p=0$)
\begin{equation}
    \rho_{\text{matter}} \propto G^{-1} a^{-3}\,,
\end{equation}
which implies a dependence of the matter mass to $G$ such that $m \propto G^{-1}$. This means either that the number of baryons is no longer conserved or that the rest energy of one individual particle depends on $G$.
Either of these options leads to questionable conclusions from a particle physics perspective. Therefore, we aim to preserve the usual conservation relation for $\D_\mu T^{\mu\nu} =0$. From a Lagrangian perspective, this means that we want to keep the usual $\sqrt{-g}$ coupling of matter and gravity $\sqrt{-g}\mathcal{L}_{\text{matter}}$, as in Ref.~\cite{Xue2015}:
\begin{equation}
\label{eq:conservation energy}
\D_\mu T^{\mu \nu}= 0 \quad\Rightarrow\quad\dot{\rho}+3H(\rho+p)=0\,.
\end{equation}

A simple way of decoupling matter conservation from $G$ and satisfying Eq.~(\ref{eq:conservation energy}) is to add a new dynamical component $S^{\mu \nu}$ to Einstein's equations. Any symmetric rank 2 tensor can be uniquely decomposed as $A u^\mu u^\nu +2q_{(\mu}u_{\nu)}+B\gamma_{\mu\nu}+\pi_{\mu\nu}$, where $A$ and $B$ are two scalar functions, $u^\mu$ is a vector field such that $u_\mu u^\mu=-1$, $q^\mu$ is a vector field transverse to $u^\mu$ ($q_\mu u^\mu=0$), $\gamma_{\mu\nu}=u_\mu u_\nu+g_{\mu\nu}$ is the transverse projector, and $\pi_{\mu\nu}$ is a symmetric transverse tensor such that $\pi_{\mu\nu}u^\mu=0$ and $\pi_{\mu\nu}\gamma^{\mu\nu}=0$. Taking $u^\mu$ as the Hubble flow and assuming isotropy and homogeneity, one has finally $q^\mu=0$ (no energy flux with respect to a Hubble observer) and $\pi_{\mu\nu}=0$ (no anisotropic pressure). Therefore, we are left with only two scalar functions, which can be rewritten as
\begin{equation}
\label{eq:tmunu}
S^{\mu \nu}=(\Phi +\Psi)u^\mu u^\nu + \Psi g^{\mu \nu}\,,
\end{equation}
with $\Phi(t)$ and $\Psi(t)$ being arbitrary functions of time.
Then, our modified equations read
\begin{equation}\label{eq:einstein}
G^{\mu \nu}= 8\pi G(t)T^{\mu \nu}+ 8\pi S^{\mu\nu} \,,
\end{equation}
where $8 \pi$ is put for convenience. The application of the Bianchi identity to Eq.~(\ref{eq:einstein}), together with~(\ref{eq:conservation energy}), then leads to
\begin{equation}
\D_{\mu}S^{\mu \nu}=-T^{\mu \nu}\partial_{\mu}G\,.
\end{equation}

The new component $S^{\mu\nu}$ is clearly not conserved if $G$ depends on time. 
Here, the only equation containing information is the temporal one ($\nu=0$) due to spatial symmetry. We can use this equation to obtain a relation between $\Phi$ and $\Psi$. Defining an effective equation of state parameter $w=\Phi/\Psi$, one gets
\begin{equation}\label{conservation for t}
    \dot{\Phi} + 3H(1+w)\Phi = -\dot{G} \rho\,.
\end{equation}

This is essentially a generalization of the energy conservation Eq.~(\ref{eq:conservation energy}) for a component coupled to the variation of $G$. Equation~(\ref{conservation for t}) shows that this component is also coupled to matter and radiation when $\dot{G}\neq0$. Conversely, the energy densities for matter and radiation, $\rho$, do not depend on $\Phi$ or $G$ directly, but relate to them through the background relation of $H$, as intended. In addition, since Eq.~(\ref{conservation for t}) is a general expression, it is also valid for a constant $G$, in which case the right-hand side becomes zero and the equation represents the conservation of energy for models of dark energy fluids uncoupled to matter and radiation.

Using the Friedmann-Lema{\^i}tre-Roberston-Walker metric with these modified Einstein equations, we directly obtain the cosmological equations in the usual manner:
\begin{align}
\label{friedmann1}
H^2 &= \frac{8 \pi G\rho}{3}+\frac{8 \pi \Phi}{3} - \frac{\kappa}{a^2}\,,\\
\frac{\ddot{a}}{a}&= -\frac{4 \pi G}{3} (\rho + 3p) -\frac{4 \pi}{3}(\Phi + 3 \Psi)\,.
\end{align}
where $\kappa$ is a constant accounting for the spatial curvature of the Universe.

\section{Selecting specific models for $\Phi$ and $\Psi$}\label{sec4}

In order to use these equations for cosmological analyses, we need to specify a function for $\Phi$. This requires making additional assumptions about the nature of this new component. To do this, let us first solve Eq.~(\ref{conservation for t}) for $\Phi$. This can be done by defining an auxiliary function $\xi(t)$ satisfying the equation $\dot{\xi}/\xi=3H(1+w)$. With this, Eq.~(\ref{conservation for t}) becomes
\begin{equation}\label{eq:Phi}
    \frac{d}{dt}(\Phi \xi)=-\dot{G} \rho \xi \,.
\end{equation}

We integrate this function with limits from $t=0$ to any time $t$. However, since we have the singularity at $a(0)=0$, $\rho$ approaches infinity at the initial point, there is a possibility that the integration will have a similar behavior at the lower boundary, which would make $\Phi$ diverge. With this in mind, we treat the lower boundary with some care:
\begin{equation}\label{eq:Phi1}
    \Phi(t) \xi(t)=\lim_{\varepsilon \to 0} \bigg(\Phi(\varepsilon) \xi(\varepsilon)  - \int_\varepsilon^t \dot{G} \rho \xi dt \bigg)\,.
\end{equation}

Let $F$ be a primitive of $\dot{G}\rho\xi$ such that $\int_\varepsilon^t \dot{G} \rho \xi dt=F(t)-F(\varepsilon)$:
\begin{equation}\label{eq:Phi2}
    \Phi(t) \xi(t)= - F(t) +\lim_{\varepsilon \to 0} \bigg(\Phi(\varepsilon) \xi(\varepsilon)  + F(\varepsilon) \bigg)\,.
\end{equation}

Then, the condition for $\Phi(t) \xi(t)$ to be finite anywhere is
\begin{equation}
    \lim_{\varepsilon \to 0} \bigg(\Phi(\varepsilon) \xi(\varepsilon) + F(\varepsilon) \bigg)=\rm{Cst.} \equiv \rm{C_1}\,.
\end{equation}
which defines a constant we call $C_1$. A similar argument exists for the auxiliary function~$\xi$. First, solving $\dot{\xi}/\xi=3H(1+w)$, we have
\begin{equation}
    \xi(t)= \lim_{\varepsilon \to 0} \Bigg[ \xi(\varepsilon) \exp\bigg(\int_{\varepsilon}^t 3 H(t)(1+w) dt \bigg) \Bigg]\,.
\end{equation}

Integrating by parts gives
\begin{equation}    \label{eq:xi de t}
    \xi(t)= \lim_{\varepsilon \to 0} \Bigg[ \xi(\varepsilon) \frac{a(t)^{3(1+w(t))}}{a(\varepsilon)^{3(1+w(\varepsilon))}} \exp\bigg(-3\int_{\varepsilon}^t  \dot{w}\,\ln(a)\, dt \bigg) \Bigg]\,.
\end{equation}

Assuming that the function $w$ is chosen with care, the expression with the exponential will converge. This leads to the condition for $\xi(t)$ to be finite:
\begin{equation}
    \lim_{\varepsilon \to 0} \Bigg[ \xi(\varepsilon) a(\varepsilon)^{-3(1+w(\varepsilon))} \Bigg] %= \rm{cst.} 
    \equiv \rm{C_2}\,.
\end{equation}

This defines a second constant $C_2$. However, the actual value of $C_2$ has no importance, since this factor cancels out in Eq.~(\ref{eq:Phi1}) and does not effect the value of $\Phi$. 

One particularly simple and interesting case is when $w$ is a constant. This leads to the simple expression
\begin{equation}\label{xi}
    \xi(t)= C_2 a(t)^{3(1+w)}\,,
\end{equation}
where we can see that the scale factor dependence of $1/\xi$ resembles that of a matter or radiation density for the appropriate values of $w$. With these, we can express $\Phi$ from Eq.~(\ref{eq:Phi2}) as
\begin{equation}\label{phi}
    \phi(t)= -\frac{F(t)}{\xi(t)}+\frac{C_1}{\xi(t)}\,.
\end{equation}

We now define  a critical energy density $\rho_c$, such that $H_0^2 = 8 \pi G_0 \rho_c/3$, and $\Omega=\rho/\rho_c=\Omega_m a^{-3} + \Omega_r a^{-4}$, where the subscript $0$ refers to the present time. Dividing by $H_0^2$, Eq.~(\ref{friedmann1}) becomes
\begin{equation}
\left( \frac{H}{H_0} \right)^2 = \frac{G(t)}{G_0}\Omega+\frac{\Phi}{G_0\rho_c} - \frac{\kappa}{a^2 H_0^2}\,.
\end{equation}

For clarity, we also define the parts of $\Phi$ in Eq.~(\ref{eq:Phi2}) coupled to radiation and matter separately, such that $F(t)/\xi(t) = G_0 \rho_m f_m(t) +G_0 \rho_r f_r(t)$, leading to the expression
\begin{align}
 \left( \frac{H}{H_0} \right)^2 = &\frac{G(a)}{ G_0 } \left(\Omega_r a^{-4} + \Omega_m a^{-3} \right) - \frac{\kappa}{a^2} + \frac{C}{\xi(a)}
 \nonumber\\ 
 &\quad-\Omega_r f_r(a) -\Omega_m f_m(a) \label{eq:friedmann}\,,
\end{align}
with $C=C_1/G_0 \rho_c$, and $\kappa$ is redefined to include the $H_0^2$ term. 

There are some important differences between this equation and the usual Friedmann-Lema\^itre equation in $\Lambda$CDM. First, there is the factor $G(a)/G_0$ in front of the usual terms for the matter and radiation contributions. This, of course, comes from the direct effect of changing $G$ on the gravitational energies of these components. Skipping $\kappa$ for the moment, we can see three additional terms that come from the energy component of $S^{\mu\nu}$, which we introduced in order to ensure the conservation of energy. The latter of these terms couple to the energy densities and the $G$ variation, while $C$ is a constant which will be discussed in more detail shortly.
As we can see from these, the actual variation of $G$ does not need to be very large, since the integral terms $f_r$ and $f_m$ can generate the more significant portion of the energy contribution as long as there is a nonzero evolution of $G$.

Finally, $\kappa$ is the usual curvature term and can be related to other parameters by evaluating Eq.~(\ref{eq:friedmann}) today ($a=1$, or $z=0$):
\begin{equation}\label{kappa}
%\kappa = \Omega_r(1-f_{r,0}) + \Omega_m (1-f_{m,0}) + \frac{C_1}{ \xi_0}   - 1  \,.
\kappa = \Omega_r(1-f_{r,0}) + \Omega_m (1-f_{m,0}) - 1  \,.
\end{equation}

At this point, we need additional assumptions in order to determine the full expression for $\Phi$.
This can be achieved most straightforwardly by choosing a relation for $\xi(a)$. In order to see what this function represents more explicitly, let us focus on Eq.~(\ref{eq:friedmann}). Since the functions $f_r$ and $f_m$ are nonzero only if $G$ evolves, they can be attributed to an extra energy contribution in the Friedmann-Lema\^itre equation arising from a variation of $G$. On the other hand, the constant $C$ produces in the Friedmann-Lema\^itre equation an energy contribution $C/\xi(a)$ that exists irrespective of whether $G$ evolves or not. As the scale factor dependence of this term is only through $\xi(a)$, the latter function essentially determines how the energy contribution of $C$ changes with the expansion of the Universe. 

Now, if $S^{\mu\nu}$ represents the reaction of spacetime to the varying gravitational constant, it makes sense to expect that the new component $\Phi$ should not change because space expands, but rather because $G$ varies. This means that, if $G$ is constant, $\Phi$ should also be unchanging, which leads to the choice $\xi(a)=\rm{Cst}$. Going back to Eq.~(\ref{xi}), this implies $w=-1$ or, equivalently, $\Phi=-\Psi$. 

With $\xi$ being constant, the $C/\xi$ term in Eq.~(\ref{eq:friedmann}) is the same as the cosmological constant $\Lambda$ in the standard picture.
However, in the present case, we have other terms in Eq.~(\ref{eq:friedmann}) that appear when $G$ varies with time, and we may not need this contribution at all.
In order to have $\Phi$ represent only the response of spacetime to the evolving $G$, we keep only the terms that depend on $G$, which means we choose $C=0$. By this choice, we get rid of the first cosmological constant problem, namely, the identification of $\Lambda$ with vacuum energy and the resulting large discrepancy with quantum field theory estimations, and we can test whether complying with cosmological observations is still possible without a cosmological constant. Therefore, we want to see if it is possible to explain the accelerated expansion with the secondary effect of the variation of $G$ instead of the vacuum energy.

For the cosmological tests we approximate $G$ with a power series expansion around $a=1$ to represent the series expansion of an unknown function:
\begin{equation}\label{G}
    G(a)=G_0\Bigg(1 + \sum_{n=1}^{\infty} b_n (1-a)^n \Bigg) \,.
\end{equation}

This expansion around $a=1$ is chosen to capture the behaviour of $G$ accurately around the low-redshift regime we investigate. When the high-redshift regime is considered, such a parametrization could be attached to any extrapolation capturing the behavior of the high-redshift regime without altering the current results at low redshift.

Then, with $\xi=\rm{Cst}$ and replacing $G$ with Eq.~(\ref{G}), $f_r$ and $f_m$ become, respectively,
\begin{align}
    f_r(a)= a^{-4}\Bigg[ b_1 \frac{a}{3} &+2b_2  \Bigg( \frac{a}{3} -\frac{a^{2}}{2} \Bigg)\nonumber\\  &+ 3b_3 \Bigg( \frac{a}{3} -a^{2} +a^{3} \Bigg)...\Bigg]\,,\label{fr}\\
    f_m(a)=a^{-3}\Bigg[b_1 \frac{a}{2} &+2b_2 \Bigg( \frac{a}{2} -a^{2} \Bigg) \nonumber\\ &+3b_3 \Bigg( \frac{a}{2} -2a^{2} - a^3 \ln a \Bigg)...\Bigg]\,.\label{fm}
\end{align}

The terms are written in a way to facilitate the comparison with the usual $\Omega_r$ and $\Omega_m$ terms in Eq.~(\ref{eq:friedmann}). This illustrates that $\Omega_r a^{-4}$ and $\Omega_m a^{-3}$ will dominate over $f_r$ and $f_m$ as $a$ gets smaller in the past.
Of course, these functions $f_r$ and $f_m$ become zero when $G$ is constant, i.e., $b_1=b_2=b_3=0$. With also $C=0$, Eq.~(\ref{eq:friedmann}) reduces to the usual Friedman-Lema{\^i}tre equations for CDM.

In the rest of the paper, we will assume a flat universe, i.e., $\kappa=0$. This allows us to determine one of the $b_i$ parameters of the expansion of $G$ in terms of the others, using Eq.~(\ref{kappa}), namely,
\begin{align}\label{b1}
 b_1 \Bigg[\frac{\Omega_m}{2} + \frac{\Omega_r}{3} \Bigg] &= \Omega_m \Bigg[1 +b_2 +\frac{9b_3}{2} \Bigg]\nonumber \\ &+\Omega_r \Bigg[ 1 +\frac{b_2}{3} - b_3 \Bigg]  - 1\,.
\end{align}

What has been done so far is to formulate a phenomenological variation of the gravitational constant within general relativity in a geometrically consistent way, also preserving the usual energy conservation. In this process, we obtained, in the Friedmann-Lema{\^i}tre equations, a cosmological constant term together with additional terms that couple to the matter and radiation components. Taking this cosmological constant term to be zero, we are left  solely with an energy contribution stemming from the coupling of matter and radiation with a variation of $G$. In the next sections, we will show that this picture is compatible with low-redshift cosmological probes to a great degree and is also able to conform to local constraints on the evolution of the gravitational constant ($\dot{G}/G$).

\section{Data and methodology}\label{sec5}
In this section, we describe the cosmological probes and the methodology used in this work. We use the usual low-redshift cosmological probes, type Ia supernovae, and baryon acoustic oscillations, together with the $\chi^2$ minimization method to constrain our model parameters. On the other hand, an investigation of the Cosmic Microwave Background measurements is left for a future work, since the calculation of the effects this variable $G$ would have on the CMB is an involved task that deserves a more through treatment. Consequently, we only consider the low-redshift regime when presenting and discussing our results.

\subsection{Type Ia supernovae}\label{sec31}
We use the SNIa measurements from the SDSS-II/SNLS3 Joint Light-Curve Analysis (JLA) dataset and its covariance matrix provided by Ref.~\cite{betoule}.
We obtain the observed distance modulus following the standardization method given by the authors:
\begin{equation}
\mu_{\text{obs}}=m-M+\alpha X-\beta C\,.
\end{equation}

In this equation, $m$, $X$, and $C$ are the observed magnitude in the B-band rest frame and the shape and color standardization parameters for the different SNIa, respectively, provided in the public dataset. The remaining parameters $\alpha$, $\beta$, and $M$ are nuisance parameters, determined together with the cosmological parameters from the fit to the data. The former two are the same for all SNIa, while the latter is the absolute magnitude in the B-band rest frame. Depending on the stellar mass of the host galaxy ($M_{\text{stellar}}$), it is given by an additional nuisance parameter $\Delta M$:
\begin{equation}
    M=
    \begin{cases}
      M', & \text{if}\ M_{\text{stellar}}<10^{10} M_\odot\,, \\
      M'+\Delta M, & \text{otherwise}\,.
    \end{cases}
  \end{equation}
  
When the gravitational strength changes due to the variation of the gravitational constant, SNIa intrinsic luminosity should also change due to the $G$ dependence of Chandrasekhar's mass, such that $L \propto M_{\text{Ch}} \propto G^{-3/2}$~\cite{gaztanaga,garciaberro,Riazuelo,Kazantzidis,mould_2014}. This modifies the observed distance modulus: If gravity was stronger in the past, for instance, supernovae would be dimmer, so their distances would actually be smaller than they appear. The required relation between the distance modulus and $G$ is directly obtained from the definition of the distance modulus. Considering that the absolute magnitude is related to luminosity via the flux, $F \propto L$ and $M=-2.5\log F(10~\text{pc})$, the distance modulus is given by
\begin{equation}
\mu_{\text{obs}}=\mu_{\text{obs},0}-\frac{15}{4}\text{log}\bigg(\frac{G}{G_0}\bigg)\,.
\end{equation}

However, there are also other approaches in the literature.
The authors of Ref.~\cite{wrightli} present an opposite relation, by using a semianalytical model to predict the SNIa light curves when the gravitational constant changes with the redshift.
Their numerical relation is converted to an approximate expression in Ref.~\cite{Sakstein2019} as $L \propto G^{1.46}$. Since there is no consensus on the precise nature of supernovae physics, as a second case we also use the distance modulus derived from this luminosity-$G$ dependence:
\begin{equation}
\mu_{\text{obs}}=\mu_{\text{obs},0}+3.65\text{log}\bigg(\frac{G}{G_0}\bigg)\,,
\end{equation}
in order to check the sensitivity of our calculations to the effect of changing $G$ on SNIa luminosities.

We compare the SNIa distance modulus to the predictions of our cosmological models using the standard definition
\begin{equation}
\mu=5 \log_{10}\left(H_0(1+z)d_M \right)\,, 
\end{equation}
with
\begin{equation}
d_M=\int_0^z \frac{\dd z'}{H(z')}
\end{equation}
being the comoving distance for a
flat space given in natural units, where $c=1$.

\subsection{Baryon acoustic oscillations}
In this work, we use a variety of isotropic and anisotropic measurements of baryon acoustic oscillations given in the literature. Isotropic observations measure the quantity $D_V/r_d$, where $r_d$ is the length of the standard ruler and $D_V$ relates to cosmology as
\begin{equation}
D_V(z)=\left( d_M^2(z) \frac{z}{H(z)} \right)^{1/3}\,.
\end{equation}

Anisotropic observations measure two quantities in the transverse and radial directions:
\begin{align}
    \theta&=\frac{r_d}{d_M}\,,\\
    \delta z_s&=r_dH(z)\,.
\end{align}

In both cases, there is a degeneracy between $H_0$ and $r_d$, so we calculate them together as a single parameter.
In this work, we assume that the variation of $G$ is small enough to not influence the BAO and use the data without modifications.

In this analysis, we consider the measurements from 6dFGS~\cite{beutler} at $z=0.106$, SDSS-MGS~\cite{Ross} at $z=0.15$, BOSS DR12~\cite{alam} at $z=0.38,\,0.51,\,0.61$, and eBOSS DR14~\cite{gilmarin} at $z=1.19,\,1.50,\,1.83$, as well as the Ly-$\alpha$ autocorrelation function~\cite{bautista} and Ly-$\alpha$-quasar cross-correlation~\cite{dumas} at $z=2.4$. We take into account the covariances for the BOSS and eBOSS measurements, we consider a correlation coefficient of $-0.38$ for the Ly-$\alpha$ forest measurements, and we assume measurements of different surveys to be uncorrelated.

Further from the peak value, the likelihoods of BAO observables diverge from a Gaussian distribution. In order to take this into account and be more conservative with our estimations, we follow the recipe in Ref.~\cite{Bassett} and replace the standard $\Delta \chi^2_G=-2\ln L_G$ likelihood expression for a Gaussian distribution with
\begin{equation}
\Delta \chi^2 = \frac{\Delta \chi^2_G}{\sqrt{1+\Delta \chi^4_G\left(\frac{S}{N}\right)^{-4}}}\,,
\end{equation}
where $S/N$ stands for the detection significance, in units of $\sigma$. We consider a detection significance of $2.4\sigma$ for 6dFGS, $2\sigma$ for SDSS-MGS, $9\sigma$ for BOSS DR12, $4\sigma$ for eBOSS DR14, and $5\sigma$ for the Ly-$\alpha$ forest.

\subsection{Determination of the parameter constraints}
Using a frequentist approach, we obtain the best-fit values for the parameters by minimizing the expression
\begin{equation}
\chi^2=(r_{\text{pred}}-r_{\text{obs}})^T C^{-1}(r_{\text{pred}}-r_{\text{obs}})\,,
\end{equation}
where $r_{\text{pred}}$ and $r_{\text{obs}}$ are the vectors that include the model prediction and the observations at each redshift, respectively, and $C$ is the covariance matrix of the observations. We add the $\chi^2$ values corresponding to each probe with the assumption that they are statistically independent. To minimize this function, we use \texttt{PYTHON}'s iminuit module\,\footnote{\url{https://pypi.org/project/iminuit/}}, an implementation of SEAL Minuit, developed at CERN~\cite{iminuit}.

\section{Results and discussion}\label{sec6}

Table~\ref{parametervalues} shows the $\chi^2$ values of our varying-$G$ model and of the standard flat $\Lambda$CDM model. The best-fit values of the parameters are also shown. We remind that $b_1$ is obtained through Eq.~(\ref{b1}). In order to obtain the uncertainty of $b_1$ we generate $10^6$ random sets of parameters from an $N$-dimensional Gaussian centered at the best fit and with the corresponding covariance matrix. For each one of these sets we derive the value of $b_1$ and compute the uncertainty from the standard deviation.

\begin{figure}[tbp]
\centering
  \includegraphics[width=\linewidth]{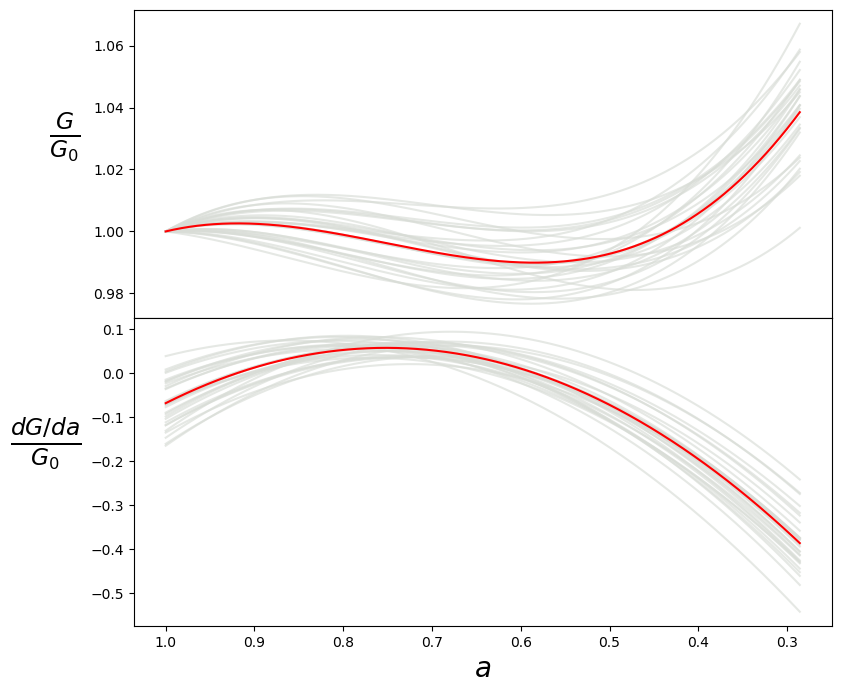}
  
\caption{Variation of $G$ and its first derivative versus the scale factor. The red line is the reconstruction using the best-fit values for the parameters. The gray lines are some sample lines with $\Delta \chi^2<1$ (see the text for details).
}\label{Gfig}
\end{figure}

\begin{figure}[tbp]
\centering
  \includegraphics[width=\linewidth]{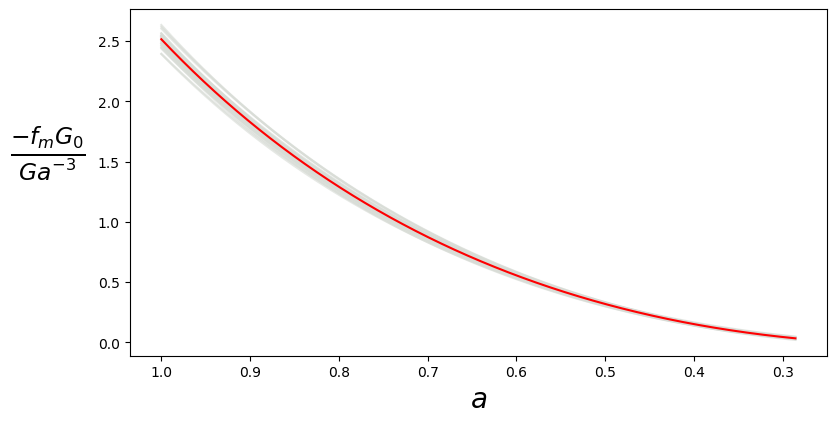}
  
\caption{Ratio of $-\Omega_m f_m$ and $\Omega_m a^{-3} G/G_0$ versus the scale factor in order to compare the contribution of the different factors driving the expansion of the Universe in the considered period. The red line is the reconstruction using the best-fit values for the parameters. The gray lines are some sample lines with $\Delta \chi^2<1$ (see the text for details). 
}\label{ffig}
\end{figure}

The reconstruction of the $G(z)$ function is shown in Fig.~\ref{Gfig}, where the red line is drawn using the best-fit values and the gray lines show sample lines with $\Delta \chi^2<1$. Again, these lines are obtained by generating random sets of parameter values from an $N$-dimensional Gaussian centered at the best fit and with the corresponding covariance matrix. The top panel in Fig.~\ref{Gfig} shows the $G(a)$ function itself, while the second panel shows the first derivative with respect to the scale factor. In both plots the functions have been normalized with respect to $G_0$, the present-day value of $G$. Constructed in a similar way, Fig.~\ref{ffig} shows the ratio between the different terms in Eq.~(\ref{eq:friedmann}) that drive the expansion at the considered epoch, $-\Omega_m f_m$ and $\Omega_m a^{-3}G/G_0$. This graph shows that the former term starts to dominate at the late stages, causing the accelerated expansion.

\begin{table*}[ht!]
  \begin{center}
      \caption{Best-fit values obtained for the parameters of the different models considered, together with the $\chi^2$ value found for each model.
      }
    \label{parametervalues}
    \begin{tabular}{c c c c c c c c} % <-- Alignments: 1st column left, 2nd middle and 3rd right, with vertical lines in between
    \hline\hline
     Model & $\chi^2$/d.o.f. & \text{$b_1$} & \text{$b_2$} & \text{$b_3$} & \textbf{$\Omega_m$}& \textbf{$\Omega_r$}&  \text{$H_0 r_d$ [km s$^{-1}$ ]} \\
      \hline
   $\Lambda$CDM  & 698.05/(756-7) &\footnotesize ... & \footnotesize ... & \footnotesize ... & \footnotesize $0.291 \pm 0.017$ & \footnotesize ($0.0 \pm 5.8) \times 10^{-3}$  & \footnotesize $(101.3 \pm 1.3) \times 10^{2}$  \\
   Varying $G$ &  697.73/(756-9) &\footnotesize $0.07 \pm 0.15$ & \footnotesize $-0.51 \pm 0.33$ & \footnotesize $0.679 \pm 0.094$ & \footnotesize $0.284 \pm 0.017$ & \footnotesize ($0.0 \pm 7.0) \times 10^{-3}$  & \footnotesize $(101.7 \pm 1.3) \times 10^{2}$ 
    \end{tabular}
  \end{center}
\end{table*}

It is clear from Table~\ref{parametervalues} that the varying $G$ model has almost the same $\chi^2$ value as the flat $\Lambda$CDM model we use for comparison. Therefore, this model is indeed capable of explaining SNIa and BAO observations without the addition of a cosmological constant. While it might seem surprising that such a small variation of the gravitational constant can result in a contribution large enough to supply most of the energy in the Universe, it is apparent from Eq.~(\ref{eq:friedmann}) and Fig.~\ref{ffig} that the major effect of $G$ on the energy balance does not come from the $G/G_0$ term. This energy instead comes from the last term on the right-hand side of this equation, $\Omega_m f_m$, which appears as a secondary effect of a time evolution of $G$. This also explains why the first-order term in the $G$ function can be so small: The contribution of $b_1$ in Eq.~(\ref{fm}) is not any larger compared to the other terms, $b_2$ and $b_3$, inside the brackets.

One of the major challenges when predicting a variation of the gravitational constant comes from the highly tight constraints measured from local observations. The review in Ref.~\cite{Uzan} compiles the different observations and gives
\begin{equation}
    \left.\frac{\dot{G}}{G} \right|_0 = (4 \pm 9) \times 10^{-13}\,\text{yr}^{-1}\,,
    \label{eq:Gdot}
\end{equation}
which is obtained from the lunar laser ranging experiment~\cite{lunarlaser}, as the tightest constraint on the variation of $G$. With the series expansion given by Eq.~(\ref{G}), the quantity $\left.\frac{\dot{G}}{G} \right|_0$ can be simply evaluated using the $b_1$ parameter:
\begin{equation}
    \frac{\dot{G}}{G}=\frac{G'(a)}{G}Ha\;\Rightarrow\;
    \left.\frac{\dot{G}}{G} \right|_0 = -b_1H_0\,,
\end{equation}
where the prime denotes the derivative with respect to the scale factor and $H_0$ is around $67-76\times 10^{-12}$ yr$^{-1}$, according to various recent measurements~\cite{Verde2019}. Looking at Table~\ref{parametervalues}, we can see that our results are compatible with
this constraint at the one-sigma level. Moreover, since our model is compatible with $b_1=0$ at one sigma, even lower values of $\frac{\dot{G}}{G}$ at $z \approx 0$ are consistent with our results.

Bounds on the variation of $G$ for earlier times also exist, such as stellar observations for low redshifts and big bang nucleosynthesis (BBN) and CMB measurements for much higher redshifts. In the former case, the constraints are much less limiting than the Solar System observations~\cite{Uzan}, and they usually assume a monotonic $G$ evolution (as in Ref.~\cite{Guenther_1998}, for instance), which is not the case in our calculations. 
In the case of BBN and CMB, the limits on $G$ evolution concern much higher redshifts than the ones our analysis considers, and, therefore, they are outside the scope of this work.

One new way of falsifying alternative gravity theories is provided by the recent observation of the GW170817 neutron star merger, as it has shown with great accuracy the gravitational wave propagation speed to be equal to the speed of light~\cite{Abbott2017}. Here, we will not provide a full mathematical demonstration but only point out that, since we do not change the geometry of spacetime, the gravitational wave propagation speed remains the same as in standard general relativity. What our approach rather does is analogous to adding background source terms, which does not affect the propagation speed. On the other hand, as shown in Ref.~\cite{Dalang2019}, modifications of the gravitational constant may cause the standard siren luminosity distance to differ from its electromagnetic counterpart.
As the capabilities of gravitational wave observatories increase, in the future this may potentially be used to probe the history of a possible $G$ evolution, but since the only available observation, GW170817, is from a very low redshift, it does not put an additional constraint to our model at the present.

Turning to the other values in Table~\ref{parametervalues}, we see that, for the varying-$G$ model, the best-fit value of $\Omega_r$ is consistent with zero, implying a very small $\rho_r$ at the present epoch, as expected. On the other hand, we obtain $\Omega_m=0.284 \pm 0.017$, which is similar to the usual value for $\Lambda$CDM. 
Therefore, we see that our model does not change the matter content drastically. The best-fit value of the $H_0 r_d$ parameter is also in agreement with the standard results.

When we repeat our analysis with the SNIa intrinsic luminosity-$G$ relation from Ref.~\cite{wrightli}, as discussed in Sec.~\ref{sec31}, we find a $\chi^2$ value slightly larger than the results discussed so far, $\chi^2=698.48$, but still perfectly compatible with the value found for $\Lambda$CDM, $\chi^2=698.05$. The values of the series expansion parameters for $G$ also change somewhat, with
\begin{align}
    b_1&=0.22 \pm 0.14\,,\\
    b_2&=-0.95 \pm 0.33\,,\\ b_3&=0.707\pm 0.083\,.
\end{align}
Most notably, $b_1$ becomes compatible with zero at two sigma instead of one. On the whole, the differences are not too drastic considering that the two supernovae luminosity models are completely opposite to each other, which leads to the conclusion that our approach does not depend heavily on the exact nature of supernovae physics. This is not surprising, since the absolute variation we predict of the gravitational constant is small in both cases, and its effect on supernovae should likewise be slight.

\section{Conclusion}\label{sec7}
In this work, we show that, when a phenomenological variation of the gravitational constant is allowed, general relativity can explain the low-redshift accelerated expansion of the Universe without a cosmological constant. 
When $G$ is taken as a time-dependent function in Einstein's field equations, the enforcement of the Bianchi identity and the usual energy conservation causes a new term to appear. This term represents the coupling between the variation of $G$ and the energy density of matter and radiation, and we determine further properties of it by requiring that it can be interpreted as a reaction of spacetime to the variation of $G$. We then test the resulting model with SNIa and BAO data.

The comparison with observational data shows that this extra term can cause the late-time accelerated expansion with a deviation less than 10\% from the current value of $G$ in the considered redshift range. We show that the required corrections to the gravitational constant are essentially on the second and third order, while the first order turns out to be small, consistent with zero within one standard deviation. This implies that the most strict bounds on a possible variation of the gravitational constant from local observations are also satisfied. 

From these results, we observe that in our varying-$G$ model the main driving force behind the accelerated expansion is not the direct effect of the gravitational constant itself but the influence of the additional term that appears because of the time dependence of $G$. While this term is dominated by the contributions of matter and radiation for higher redshifts, it starts to supply most of the energy in the Universe during the late stages and thereby facilitates the acceleration. As a result, we see that this model can explain the late-stage accelerated expansion of the Universe without a cosmological constant and without requiring a large impact on the small-scale gravitational processes.

Finally, let us mention that varying-$G$ models also have a broader potential in explaining other cosmological tensions. As an example, it has been shown recently that a cosmological Brans-Dicke model with a cosmological constant can alleviate the tension on the Hubble constant $H_0$~\cite{Sola_2019}. While this discussion is outside the scope of this paper, it shows the potential of considering varying-$G$ models as interesting alternatives to the standard general relativity.

\bibliography{apssamp}% Produces the bibliography via BibTeX.

\end{document}